\RequirePackage{fix-cm}
\documentclass[twocolumn,epjc3]{svjour3}  
\smartqed  
\RequirePackage{graphicx}
\RequirePackage{latexsym}
\usepackage{amsmath}
\usepackage{amssymb}
\RequirePackage[numbers,sort&compress]{natbib}
\RequirePackage[colorlinks,citecolor=blue,urlcolor=blue,linkcolor=blue]{hyperref}
\usepackage[colorlinks]{hyperref}

\journalname{Eur. Phys. J. C}
\begin{document}

\title{Radiopurity of a kg-scale PbWO$_4$ cryogenic detector produced from archaeological Pb for the RES-NOVA experiment}


\author{
The \mbox{\protect{\sc{RES-NOVA}}} group of interest
        \and  \\[2mm]
J.W.~Beeman \thanksref{LBL}
\and
G.~Benato \thanksref{LNGS}
\and
C.~Bucci \thanksref{LNGS}
\and
L.~Canonica \thanksref{MPI}
\and
P.~Carniti \thanksref{UNIMIB, INFN-MIB}
\and
E.~Celi \thanksref{LNGS,GSSI}
\and
M.~Clemenza \thanksref{INFN-MIB}
\and
A.~D'Addabbo \thanksref{LNGS}
\and
F.A.~Danevich \thanksref{INR}
\and
S.~Di Domizio \thanksref{Ge}
\and
S.~Di~Lorenzo \thanksref{LNGS}
\and
O.M.~Dubovik \thanksref{ISMA}
\and
N.~Ferreiro Iachellini \thanksref{MPI}
\and
F.~Ferroni \thanksref{GSSI,Roma1}
\and
E.~Fiorini \thanksref{UNIMIB, INFN-MIB}
\and
S.~Fu \thanksref{LNGS}
\and
A.~Garai \thanksref{MPI}
\and
S.~Ghislandi \thanksref{LNGS,GSSI}
\and
L.~Gironi \thanksref{UNIMIB, INFN-MIB}
\and
P.~Gorla \thanksref{LNGS}
\and
C.~Gotti \thanksref{UNIMIB, INFN-MIB}
\and
P.V.~Guillaumon \thanksref{LNGS}
\and
D.L.~Helis \thanksref{LNGS,GSSI}
\and
G.P.~Kovtun \thanksref{NSC}
\and
M.~Mancuso \thanksref{MPI}
\and
L.~Marini \thanksref{LNGS,GSSI}
\and
M.~Olmi \thanksref{LNGS}
\and
L.~Pagnanini \thanksref{LNGS}
\and
L.~Pattavina \thanksref{e1,LNGS,TUM}
\and
G.~Pessina \thanksref{INFN-MIB}
\and
F.~Petricca \thanksref{MPI}
\and
S.~Pirro \thanksref{LNGS}
\and
S.~Pozzi \thanksref{UNIMIB, INFN-MIB}
\and
A.~Puiu \thanksref{e1, LNGS, GSSI}
\and
S.~Quitadamo \thanksref{LNGS,GSSI}
\and
J.~Rothe \thanksref{TUM}
\and
A.P.~Scherban \thanksref{NSC}
\and
S.~Sch{\"o}nert \thanksref{TUM}
\and
D.A.~Solopikhin \thanksref{NSC}
\and
R.~Strauss \thanksref{TUM}
\and
E.~Tarabini \thanksref{INFN-MIB}
\and
V.I.~Tretyak \thanksref{INR}
\and
I.A.~Tupitsyna \thanksref{ISMA}
\and
V.~Wagner \thanksref{TUM}
}

\thankstext{e1}{Corresponding authors: luca.pattavina@lngs.infn.it \\ \hangindent=3.5cm  andrei.puiu@gssi.it}


\institute{Lawrence Berkeley National Laboratory, Berkeley, California 94720, USA  \label{LBL}
\and
INFN Laboratori Nazionali del Gran Sasso, Via G. Acitelli 22, I-67100 Assergi, Italy \label{LNGS}
\and
Max-Planck-Institut f{\"u}r Physik,  F{\"o}hringer Ring 6, DE-80805 M{\"u}nchen, Germany \label{MPI}
\and
Dipartimento di Fisica, Universit\`a di Milano - Bicocca, Piazza della Scienza 3, I-20126 Milano, Italy \label{UNIMIB}
\and
INFN Sezione di Milano - Bicocca, Piazza della Scienza 3, I-20126 Milano, Italy \label{INFN-MIB}
\and
Gran Sasso Science Institute,  Viale F. Crispi 7, I-67100 L’Aquila, Italy \label{GSSI}
\and
Institute for Nuclear Research of NASU, 03028 Kyiv,	Ukraine \label{INR}
\and
INFN Sezione di Genova and Universit\`a di Genova, Via Dodecaneso 33, I-16146 Genova, Italy \label{Ge}
\and
Institute of Scintillation Materials of NASU, 61072  Kharkiv, Ukraine \label{ISMA}
\and
INFN Sezione di Roma-1, P.le Aldo Moro 2, I-00185 Roma, Italy \label{Roma1}
\and
National Science Center 'Kharkiv Institute of Physics and Technology', 61108 Kharkiv, Ukraine \label{NSC}
\and
Physik-Department and Excellence Cluster Origins, Technische Universit{\"a}t M{\"u}nchen, James-Franck-Stra{\ss}e 1, DE-85747 Garching, Germany \label{TUM}
}

\date{Received: date / Accepted: date}

\maketitle

\begin{abstract}
RES-NOVA is a newly proposed experiment for the detection of neutrinos from astrophysical sources, mainly Supernovae, using an array of cryogenic detectors made of PbWO$_4$ crystals produced from archaeological Pb. This unconventional material, characterized by intrinsic high radiopurity, enables to achieve low-background levels in the region of interest for the neutrino detection via Coherent Elastic neutrino-Nucleus Scattering (CE$\nu$NS). This signal lies at the detector energy threshold, \textit{O}(1~keV), and it is expected to be hidden by naturally occurring radioactive contaminants of the crystal absorber. Here, we present the results of a radiopurity assay on a 0.84~kg PbWO$_4$ crystal produced from archaeological Pb operated as a cryogenic detector. The crystal internal radioactive contaminations are: $^{232}$Th $<$40~$\mu$Bq/kg, $^{238}$U $<$30~$\mu$Bq/kg, $^{226}$Ra 1.3~mBq/kg and $^{210}$Pb 22.5~mBq/kg. We present also a background projection for the final experiment and possible mitigation strategies for further background suppression. The achieved results demonstrate the feasibility of realizing this new class of detectors.

\end{abstract}

\section{Introduction}
\label{intro}

In 2017 the physics community entered the era of Multi-Messenger Astronomy (MMA). For the first time, we were able to identify in the sky two astrophysical sources, emitting with a strong time correlation, electromagnetic radiations (EMR) and neutrinos (TXS 0506+056)~\cite{TXS_0506_056} and gravitational waves (GW) together with EMR (GW170817)~\cite{GW170817}. In this overall context, the missing part is the simultaneous detection of the three components. This is currently considered one of the greatest discovery of MMA, but also more in general of Modern Physics.

Among the high-energy cosmic events which might lead to a threefold multi-messenger observation, Supernovae (SNe) have one of the most exciting prospects. In fact, during this event a massive star ($>8~M_\odot$) undergoes gravitational collapse, and whenever this is successful the star explodes and neutrinos are emitted copiously ($\mathcal{O}(10^{57})$) together with GWs. An electromagnetic transient is also expected to be observed, at a later time, when stellar dusts do not obscure the line of sight. For SN1987A~\cite{PhysRevLett.58.1490}, the EM counterpart of the SN emission was detected about 2-3~hours after the arrival of the neutrino burst~\cite{1987Nature}.

The combination of observing neutrinos and GWs will provide a direct insight into SNe through different points of view. 
Neutrinos and GWs provide exclusive information into the explosion dynamics in real-time, granting access to the inner core of the envelope. In particular, a star is largely transparent to neutrinos, except the region within ~50 km of the centre of the proto-neutron star. Thus, neutrinos are direct probes of the central engine of core-collapse-SNe and carry the imprints of the nuclear and astrophysical processes occurring during the explosions.
State-of-the-art neutrino detectors will allow to extract a variety of information from this event, by characterizing the energy and time distributions of the neutrino emission, as well as the neutrino flavor partition~\cite{Mirizzi:2015eza}.

The recent technological advancements of cryogenic detectors in neutrinoless double-beta decay~\cite{ALDUINO20199,Azzolini:2018tum,Cupid-Mo,Fink:2020jts,Alenkov:2019jis} and Dark Matter (DM)~\cite{Abdelhameed:2019hmk,Strauss:2017cam} experiments opened new opportunities for the employment of such an advanced technique also for the detection of neutrinos from artificial~\cite{Strauss:2017cuu, Billard:2016giu, Aliane:2020dbc} and astrophysical~\cite{Pattavina:2020cqc} sources.
In this manu\-script, we present a detailed background study of a lead tungstate (PbWO$_4$) prototype cryogenic detector made from archaeological Pb, for the detection of SN neutrinos via Coherent elastic neutrino-Nucleus Scattering~\cite{Freedman:1973yd}. The detector represents the first kg-scale proof-of-principle detector for the technology proposed by the RES-NOVA project~\cite{Pattavina:2020cqc, RES-NOVA:2021gqp}.

In the first section of this manuscript, we discuss the detection channel used by RES-NOVA. In the second and third sections, the detector working principle and the basic procedures for the production of high-purity PbWO$_4$ crystal from archaeological Pb are introduced. Then, we present the results and performance of the cryogenic detector, as well as some projections for the full RES-NOVA experiments. Finally, an outlook for further improvements on the detector radiopurity level are discussed.

\section{Supernova neutrino detection via Coherent Elastic neutrino-Nucleus Scattering}

\begin{figure}[t]
\centering
\includegraphics[width=0.49\textwidth]{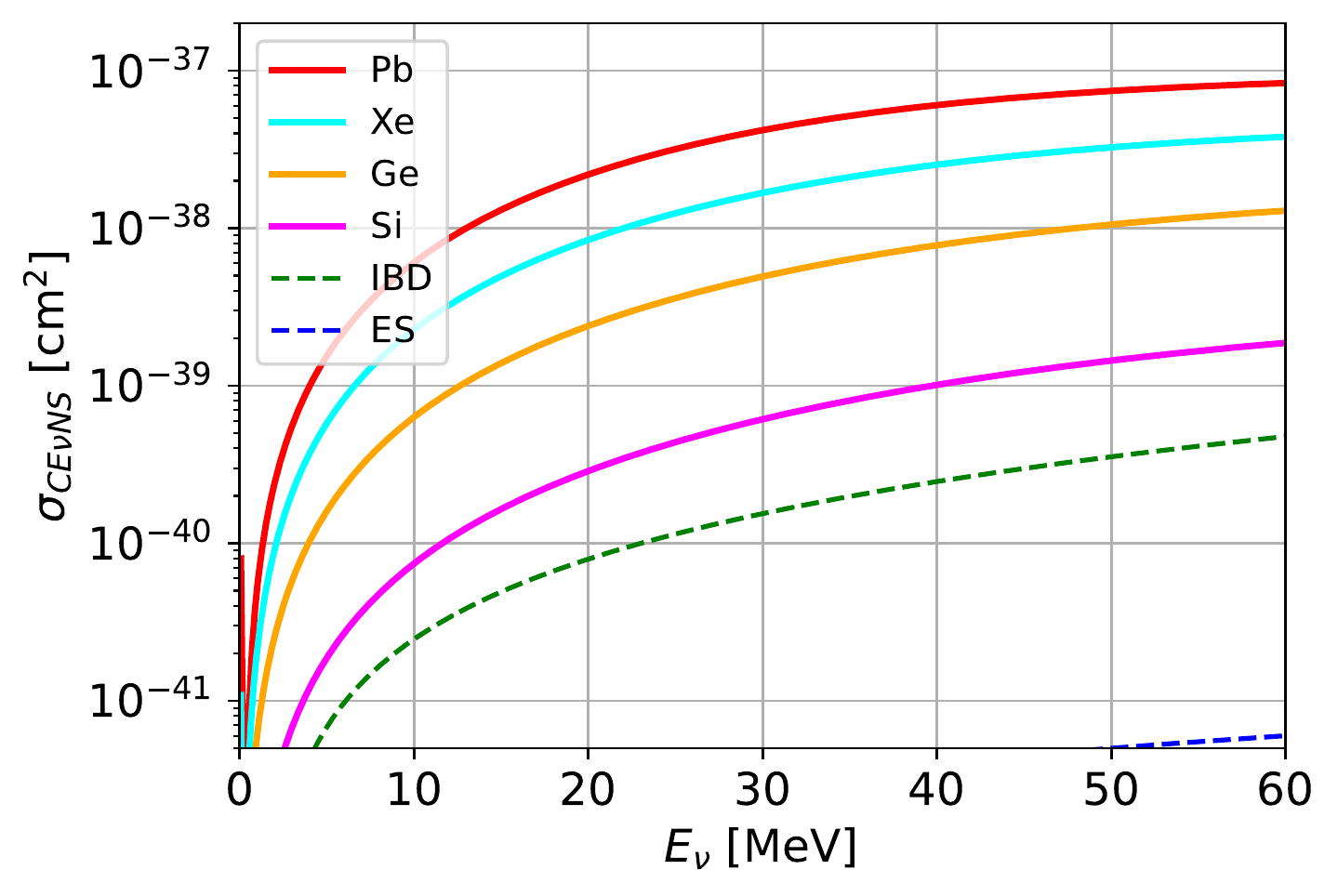}
\caption{Coherent elastic neutrino-nucleus scattering
(CE$\nu$NS) cross-sections as a function of the energy of the incoming neutrino for different target nuclei. The cross-sections for inverse-beta decay (IBD) and neutrino elastic scattering on electrons (ES) are also shown for the sake of comparison.}
\label{fig:xsection} 
\end{figure}

In the field of neutrino detectors, among the various available detection channels~\cite{SNEWS:2020tbu}, one which raised great interest in both experimentalists and theoreticians community, is Coherent Elastic neutrino-Nucleus Scattering (CE$\nu$NS)~\cite{Freedman:1973yd}. This channel was discovered in 2017~\cite{Akimov:2017ade} and it offers a wealth of applications in neutrino physics thanks to its neutral current nature and the high cross-section, which is 3-to-4 orders of magnitude higher than neutrino-electron scattering and inverse-beta decay (see Fig.~\ref{fig:xsection}). In this process, the lack of a kinematic threshold can also be exploited for detecting low energy neutrinos from nuclear reactors, the Earth's crust (i.e. geo-neutrinos) or neutrinos from the Sun and other stars.

CE$\nu$NS is a Standard Model process where a $Z^0$ is exchanged between the target nucleus and the incoming neutrino. This is the same process which is thought to be responsible for core-collapse SN explosions~\cite{Freedman:1977xn}. The cross-section for a spin-zero nucleus, neglecting radiative corrections~\cite{Prospects_Scholberg}, is given by:
\begin{equation}
\label{eq:xsec}
\frac{d\sigma}{d E_R} = \frac{G^2_F m_N}{8 \pi (\hslash c )^4} \;Q_W^2 \left( 2- \frac{E_R m_N}{E^2} \right) \cdot F^2(2E_R m_N) 
\end{equation}

\noindent where $G_F$ is the Fermi coupling constant, $m_N$ it is the mass of the target nucleus, $Q_W$ the weak nuclear charge, while $E$ is the energy of the incoming neutrino and $E_R$ the recoil energy of the nucleus. The last term of the equation, $F$, is the elastic nuclear form factor, which describes the spatial and weak charge distributions of neutrons (i.e. radius and skin) inside the target nucleus~\cite{Amanik:2009zz}. For small momentum transfers $F=1$.

The expected signal induced in a detector by a SN neutrino interacting via CE$\nu$NS is a low energy nuclear recoil in the few of keV range. A 20~MeV neutrino produces a nuclear recoil on a Pb nucleus of about 1~keV and, depending on the SN model, neutrinos have  energies of several tens of MeV~\cite{Mirizzi:2015eza}. This type of signature is similar to the one produced by DM particles scattering off a target nucleus. For this reason, DM detectors are also proposed to be exploited for the detection of SN neutrinos~\cite{Raj:2019sci}. In particular, the ones based on the Xe and Ar-liquid time projection chamber technologies
~\cite{Lang:2016zhv, Agnes:2020pbw, Horowitz:2003cz} are among the best candidate. However, while they were successful in scaling-up detectors to multi-ton scales~\cite{Schumann:2019eaa}, they are affected by limited energy resolutions at threshold (e.g. limited photon and electron statistics) and by a fixed number of available target materials (i.e. noble-liquids).

\section{Archaeological Pb-based cryogenic detectors}

The cryogenic calorimetric technique~\cite{Pirro:2017ecr} is an advanced and flexible experimental approach, which allows achieving extremely low energy threshold and high-resolution, thus enhancing the sensitivity to SN neutrino signals. At the same time, it offers the possibility to adopt a variety of target materials while preserving the same advanced performance.
RES-NOVA is a newly proposed experiment~\cite{Pattavina:2020cqc} aiming at operating PbWO$_4$ crystals made of archaeological Pb ($^{arch}$PbWO$_4$) as cryogenic detectors, for the detection of neutrinos from astrophysical sources. Pb is the only element of the periodic table that offers simultaneously the highest CE$\nu$NS cross-section (see Fig.~\ref{fig:xsection}), thanks to the high neutron number ($N$=82), and the longest nuclear stability ($\tau_{1/2} >10^{20}$~y~\cite{Beeman:2012wz}) for low-backgrounds. In addition, the employment of archaeological Pb secures an ultra-low background level in the region of interest, given the excellent radiopurity level of this valuable material~\cite{Pattavina:2019pxw}. These aspects together with the high density of PbWO$_4$ enables the realization of neutrino telescopes with reduced dimensions when compared to the conventional ones (e.g. SuperKamiokande, SNO, JUNO)~\cite{SNEWS:2020tbu}. RES-NOVA is planning to operate a detector array with a total active volume of (60~cm)$^3$, more than a factor 10$^3$ smaller than other neutrino detectors. Even with such small experimental volume, the entire Milky Way can be probed for SNe with 5$\sigma$ sensitivity~\cite{RES-NOVA:2021gqp}. Reconstruction of the main SN parameters can also be performed, achieving a precision comparable to the one of currently running experiments~\cite{Pattavina:2020cqc}.

The signal induced by SN neutrino events in RES-NOVA are nuclear recoils of few keV of energy~\cite{Pattavina:2020cqc}. Only the background level attainable with archaeological Pb enables a significant detection of SN signals. Particle interactions in the detector may mimic the signal under investigation, for this reason, all background sources must be minimized. The background in the region of interest (RoI) must be orders of magnitude lower than the neutrino signal. Cosmic-ray interactions and natural radioactivity, namely $^{238}$U and $^{232}$Th decay chains, are responsible for the background. RES-NOVA will be installed in a deep underground laboratory, as the Gran Sasso (LNGS, Italy), where the overburden of 3600~m w.e. will suppress the cosmic-ray flux by 6 orders of magnitude. The main background source in PbWO$_4$ crystals is expected to come from Pb itself. Commercial low-background Pb can not be used for RES-NOVA, due to the intrinsic overwhelming concentration of $^{210}$Pb~\cite{Pattavina:2019pxw}. This problem is overcome when archaeological Pb is used for the crystal production~\footnote{Lead-210 has a half-life of 22.3~y and it decays $\beta^-$ with a $Q$-value of 63~keV.}. However, the concentration of other dangerous nuclides, as $^{238}$U and $^{232}$Th, must also be reduced. In \cite{RES-NOVA:2021gqp}, we presented the first background budget of the RES-NOVA experiment. There, we demonstrated that the crystal radioactive contaminations are the main responsible for the background in the RoI. In this respect, archaeological Pb plays a crucial role in the operation of low background PbWO$_4$ detectors.

Recently, the first small scale proof-of-principle of the RES-NOVA technology was set into operation~\cite{Iachellini:2021rmh} in an above-ground unshielded set-up. A 16~g $^{arch}$PbWO$_4$ was equipped with a Transition Edge Sensor thermometer (TES), of the same type as the ones used for DM searches by the CRESST experiment~\cite{Abdelhameed:2020opm}, to test its performance as a low temperature calorimeter at low energies. The sensor design was not optimized for achieving ultra-low energy threshold. Despite the sensor was not optimised for an ultra-low energy threshold, it performed well enough to achieve a 300~eV threshold. This was the first milestone of the RES-NOVA project. However, the intrinsic radiopurity of the crystal was not assessed due to the high detector counting rate induced by cosmic-rays, and the small detector mass.
In this work, we assess the radioactive contaminations of a kg-scale $^{arch}$PbWO$_4$ cryogenic detector for the first time.

\section{Crystal production}

Lead is an excellent material both for passive shield for low-background detectors~\cite{Heusser:1995wd} 
and for production of PbWO$_4$ crystals to be used as light guides in scintillation rare decays experiments \cite{Belli:2016, Belli:2020}. 
PbWO$_4$ crystal scintillators, having a rather poor scintillation efficiency at room temperature, are quite efficient scintillators at low temperatures \cite{Danevich:2010}. However, ordinary Pb 
contains the radioactive isotope $^{210}$Pb with an activity on the level of tens -- thousands Bq/kg, which is unacceptable for low-background experiments. 
$^{210}$Pb is one of the daughters in the $^{238}$U decay chain, with contribution also from radioactive $^{222}$Rn gas present in the atmosphere. Once Pb is extracted from ores (galena PbS, anglesite PbSO$_4$, cerussite PbCO$_3$), the content of $^{210}$Pb decreases with time. 
Taking into account the half-life of $^{210}$Pb, the radioactivity of Pb smelted hundreds years ago is expected to be negligible~\cite{Alessandrello:1991, Alessandrello:1993}.

The $^{arch}$PbWO$_4$ crystal scintillator, studied in the present work, was developed from the archaeological Pb, obtained from a Greek ship sunk in the Black Sea in the 1$^{st}$ century B.C.~\cite{Danevich:2009}. 
To obtain a high-quality PbWO$_4$ crystal scintillator, the PbO used for the PbWO$_4$ crystal production should be both radiopure and uncontaminated in terms of other chemical elements. A complex refining scheme was developed to purify the archaeological Pb. The purification procedure included melting with filtration (to remove oxides, macro-, and micro-inclusions), distillation with vapor condensation into the liquid phase (removal of non-volatile impurities), and high-temperature heating to remove volatile impurities \cite{Boiko:2011}. 
The purification process reduced the concentrations of impurities, in particular Cu, Ag, and Sb, by a factor $\sim$510, $\sim$90 and $\sim$400, down to a level of $\leq$0.3~ppm, $\leq$0.7~ppm 
and $\sim$0.6~ppm, respectively. As a result, the purity grade of the archaeological Pb was improved to 99.9996\%.

The charge for the $^{arch}$PbWO$_4$ crystal growth was obtained by the method of high-temperature solid-phase synthesis from oxides $^{arch}$PbO and WO$_3$. 
The content of the main impurity elements in the WO$_3$ used for the crystal production was at the level of $<$1~ppm - 0.2~ppm. As a first step, lead nitrate was obtained by dissolution of the metal in a 20\% solution of nitric acid. The solution containing lead nitrate was then neutralized with gaseous ammonia, resulting in a precipitate of lead oxide PbO$\cdot$nH$_2$O. The stoichiometric lead oxide PbO was obtained by annealing of the PbO$\cdot$nH$_2$O compound at 600$^{\circ}$ for 4~h. The synthesis of the $^{arch}$PbWO$_4$ powder was carried out by stepwise heating and holding the mixture of starting oxides in air at temperatures from 250$^{\circ}$ to 800$^{\circ}$ with periodic grinding of the mixture in a ball mill. The phase composition of the powder was monitored by X-ray phase analysis.

\begin{figure}[t]
\centering
\includegraphics[width=0.4\textwidth]{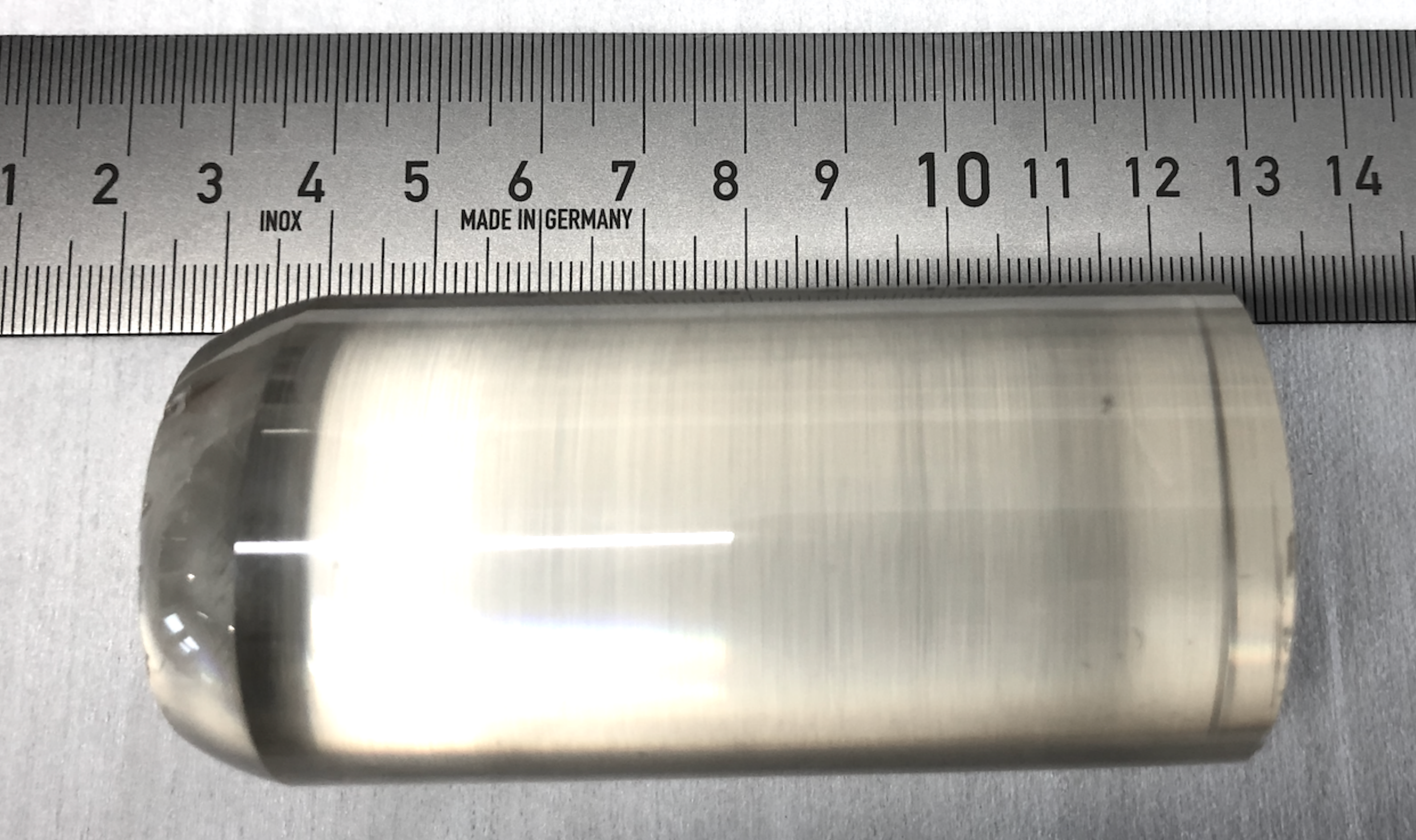}
\caption{PbWO$_4$ crystal scintillator produced from archaeological Pb. The crystal has a mass of 0.84~kg, a diameter of 40~mm and a height of 83~mm.}
\label{fig:xtal} 
\end{figure}

The $^{arch}$PbWO$_4$ crystal boule was grown by the Czochralski method by using a ``Crystal-607'' growing set-up in a platinum crucible with diameter 80~mm and 80~mm height. The crystal was pulled with a rate of $2-5$ mm per hour, and a rotation rate $10-25$ rotations per minute. From the final boule a crystal with sizes 40~mm in diameter and 83~mm in length with a shape close to a logarithmic spiral was produced~\footnote{The crystal was first used as light guide in different experiments~\cite{Belli:2016,Belli:2020}.} (mass of 0.84~kg), see Fig.~\ref{fig:xtal}. During the manufacturing process, the crystal acquired a noticeable color caused by ultraviolet radiation present in daylight, which was then removed by annealing of the crystal in an air atmosphere.

\section{Operation of a kg-scale $^{arch}$PbWO$_4$ cryogenic detector}

The $^{arch}$PbWO$_4$ crystal was housed in a highly pure Cu structure, similar to one of the following works~\cite{Cardani:2013dia, Casali:2013zzr, Pattavina:2015jxe, Artusa:2016mat}. The crystal was held in position by bronze and PTFE supports, anchored to a Cu structure. The crystal was equipped with a Ge-NTD thermistor, of the same type as the one developed for the CUORE, CUPID-0 and CUPID-Mo experiments~\cite{Pirro:2017ecr}. This sensors measures the temperature increase caused by any energy deposition in the crystal. This was also instrumented with a Si Joule heater used for the offline detector gain stabilization.
We used a Ge-NTD thermal sensor for studying the crystal radiopurity. This choice is driven by the large dynamic range (from few keV up to some MeV) and high resolution of the thermometer.

The signals induced in the Ge-NTD were amplified by means of JFET input differential voltage sensitive operated amplifier~\cite{JFET} and fed into an 18 bit NI-6284 PXI ADC unit. A 6-pole low-pass Bessel filter was placed before the ADC, with a roll-off rate of 120~dB/decade, to prevent aliasing effects on the acquired signal~\cite{ARNABOLDI2010327}. The filter cut-off frequency was set at 15~Hz. The trigger was software generated on the detector and, when it fired, 5~s long waveforms, sampled at 2~kHz, were saved on disk~\cite{Domizio_2018}. The detector was operated in its optimal working point~\cite{Azzolini:2018tum}, where the signal-to-noise ratio was maximized. An offline Optimum Filter algorithm is also applied to further enhance it~\cite{Gatti:1986cw, Azzolini:2018yye}.

The experimental set-up was sited in the underground Hall A of the Laboratori Nazionali del Gran Sasso of INFN (Italy). There, we can benefit from the rock overburden for suppressing the interaction of cosmic-rays in the experimental set-up. The detector was installed in an Oxford 1000 $^3$He/$^4$He dilution refrigerator, the very same system used for the operation of the CUORICINO~\cite{Andreotti:2010vj}, CUORE-0~\cite{Alfonso:2015wka} and CUPID-0~\cite{Azzolini:2018tum} detectors.

The main goal of the measurement was to investigate the crystal radiopurity. In particular we were interested in identifying and quantifying the concentration of radioactive contaminants inside the $^{arch}$PbWO$_4$ crystal. The largest and most relevant contributions are expected to come from radionuclides of the natural primordial decay chains of $^{232}$Th and $^{238}$U. The RoI for these studies lies at the MeV scale. At these energies, it is convenient to study the internal radioactive contaminations of the crystal, given the high signal-to-background ratio. In fact, signals at these scales are mostly produced by $\alpha$ decaying radioactive impurities in the detector bulk. The contribution from radioactive surface contaminations is expected to be negligible~\cite{Clemenza:2011zz} and can be clearly identified due to the different energy signature. In fact, for bulk contaminations the entire energy of the decay ($\alpha$ + nuclear recoil) is detected, while for surface contaminations a fraction of the decay energy is lost, due to the escaping of one of the decay products.

The detector response was energy calibrated by means of external high-energy $\gamma$-sources (i.e. $^{232}$Th), placed in between the Pb shielding of the set-up and the cryostat vacuum vessels. The source was deployed during the entire data-taking next to the experimental set-up. Crystal internal radioactive contaminations were also used for calibrating the detector in our RoI.

\begin{figure}[h]
\centering
\includegraphics[width=0.49\textwidth]{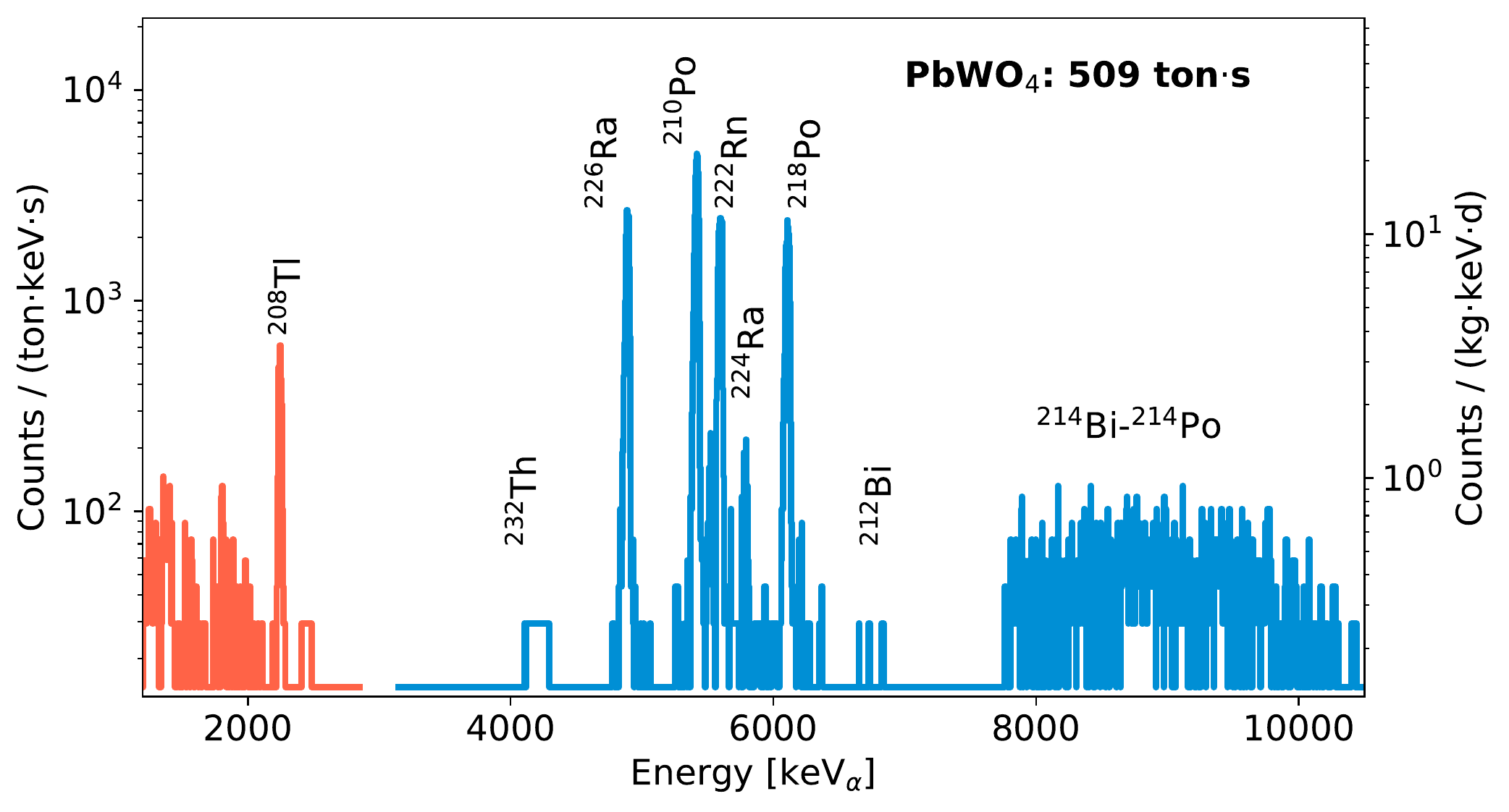}
\caption{Final detector energy spectrum of a 0.84~kg $^{arch}$PbWO$_4$ crystal operated as cryogenic detector. The total collected statistics amounts to 509~ton$\cdot$s (5.9~kg$\cdot$d). The energy spectrum is zoomed in the higher energy region of the events. The blue part of the spectrum highlights $\alpha$ induced events, while the red one $e^-$/$\gamma$ events.}
\label{fig:spectrum} 
\end{figure}

The final energy spectrum is obtained by events that passed global and event-based selection cuts. The global ones concern the identification of time period with high-noise level or detector instabilities, while the event-based ones are connected with the rejection of non-particle events (e.g. electrical spikes, disturbances). The result of this data analysis is shown in Fig.~\ref{fig:spectrum}, where the total detector energy spectrum, zoomed in the RoI is shown.

PbWO$_4$ is a highly efficient scintillator also at low temperatures~\cite{Lecoq:1994yr}. The scintillation light produced by particle interactions in the absorber can be exploited for particle identification and background suppression~\cite{Beeman:2012wz}. However, in our analysis we did not exploit this additional read-out channel given that the expected background induced by $e^-$/$\gamma$ interactions in the RoI for our \textit{$\alpha$-analysis} ($>3$~MeV) is negligible.

\section{Crystal radiopurity studies}
We acquired a total statistic of 509~ton$\cdot$s (5.9~kg$\cdot$d). We adopt this measuring units for the detector exposure because neutrino signals develops on the time scale of few second (about 10~s), and the required detectors should have masses of the order of few tons (RES-NOVA detector mass will be 1.8~tons).

In Fig.~\ref{fig:spectrum}, the final detector energy spectrum zoomed in the higher energy region is shown. 
 The main peaks visible at energies above 3~MeV are ascribed to crystal internal $\alpha$ decaying contaminations, while below 3~MeV, $e^-$/$\gamma$ events mostly induced by the external $\gamma$-source are visible. The peak at around 2.2~MeV is the characteristic $^{208}$Tl $\gamma$ emission, which is expected to be observed at 2.6~MeV. This shift in the energy calibration is caused by an energy quenching between $\alpha$s and $e^-$/$\gamma$ events. In fact, the spectrum is energy calibrated using only $\alpha$ lines. We can estimate a quenching factor for $e^-$/$\gamma$ interactions (QF) of (14.2$\pm$0.9)\%, which is slightly lower than what observed in~\cite{Beeman:2012wz}, 20\% at 2.5~MeV. Two broad distributions at low and high energies can be identified in the energy spectrum: the first one is produced by Compton interactions of the $^{208}$Tl external high-energy $\gamma$ source. The second one is caused by pile-up events of radionuclides of the $^{238}$U decay chain, namely $^{214}$Bi-$^{214}$Po. The time resolution of the thermal sensor does not allow resolving the two events, due to the very short half-life of $^{214}$Po, which is $\tau_{1/2}$=0.16~ms~\cite{Borexino:2012uda}.
 
 The evaluation of the concentration of radionuclides inside the crystal was performed following the same procedure described in~\cite{LYS}. The results of the analysis are shown in Tab.~\ref{tab:radio}. 

\begin{figure}[h]
\centering
\includegraphics[width=0.49\textwidth]{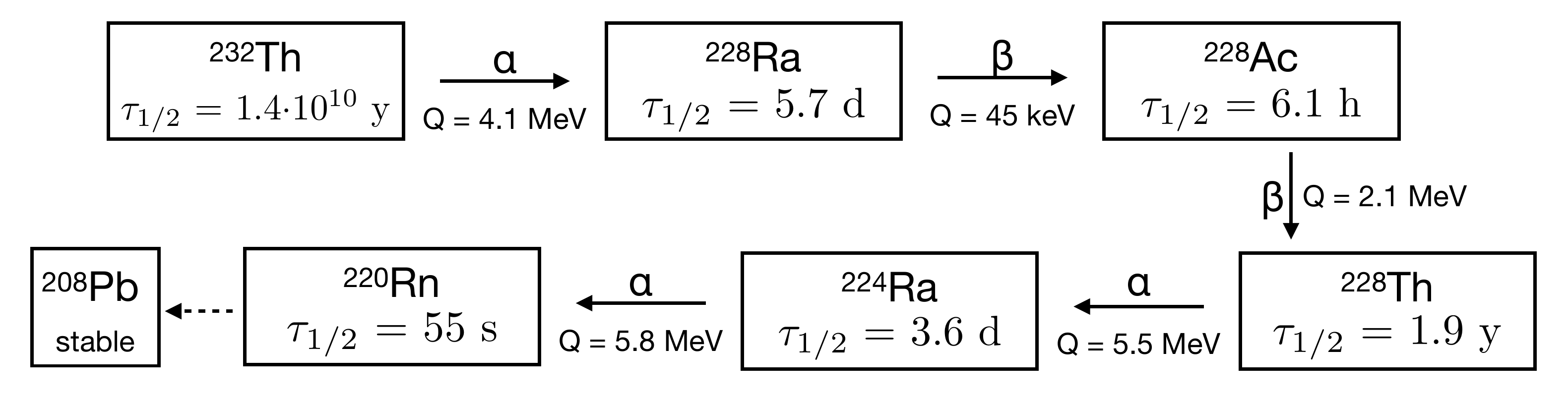}
\caption{Simplified scheme of the $^{232}$Th decay chain. The nuclides of the upper part of the chain are shown, together with their half-life, Q-value of the decay and  decay mode.}
\label{fig:Th232decay} 
\end{figure}

The crystal features a rather small internal contamination of primordial $^{232}$Th, which is not in secular equilibrium (i.e. different activities) with other nuclides of the decay chain, namely $^{228}$Ra and $^{228}$Th. The breaking is usually caused by the different chemical properties of Ra and Th, and by the production procedure for the crystal growth. This aspect is particularly relevant, because as shown in~\cite{RES-NOVA:2021gqp, Strauss:2014aqw} $^{228}$Ac is a dangerous background source for low-energy experiments, having a low energy $\beta$-decaying transition. For the sake of clarity, in Fig.~\ref{fig:Th232decay} a sketch of the $^{232}$Th decay chain scheme is shown.
There, we can identify three nuclides where the secular equilibrium might be broken, having very different half-lives with respect to their daughter nuclides: $^{232}$Th, $^{228}$Ra and $^{228}$Th. In our detector, we can clearly identify the presence of $^{232}$Th and $^{228}$Th (through the detection of $^{224}$Ra), while we can not make any robust statement about $^{228}$Ra/$^{228}$Ac. The detection of these nuclides is made arduous due to the fact that both decay $\beta$, thus producing a continuous signature in the energy spectrum. They have energy transitions falling in the region of the energy spectrum below 2.6~MeV, where the background level is higher. On the basis of these assessments, two assumptions can be made about $^{228}$Ra/$^{228}$Ac activity: i) $^{228}$Ra is in secular equilibrium with $^{232}$Th, and thus $^{228}$Ac has an activity of  $<$0.04~mBq/kg, or ii) $^{228}$Ra is in secular equilibrium with $^{228}$Th, and thus $^{228}$Ac has an activity of about 0.80~mBq/kg. These two hypotheses, as will be shown later in the manuscript, may have a different impact in the background projection of the RES-NOVA experiment. However, it is worth mentioning that from chemical arguments ($^{232}$Th and $^{228}$Th have the same chemical properties) and considering that the crystal was produced more than 2$\cdot \tau^{^{228}Th}_{1/2}$ ago, the second assumption is more likely to be valid. Future measurements with high sensitivity to low energy $e^-$/$\gamma$ emissions (e.g. using a HP-Ge detector) will address this issue. 

Radionuclides from the $^{238}$U decay chains can also be identified, in particular the sub-decay chain that goes from $^{226}$Ra to $^{210}$Po. All the elements of this chain are in secular equilibrium among them, except for $^{210}$Po, which has a higher concentration: 22.50$\pm$0.49~mBq/kg. This nuclide is assumed to be in equilibrium with the parent $^{210}$Pb, given that the crystal was produced more than 5$\cdot \tau^{^{210}Po}_{1/2}$ ago (September 2010). For this reason, we can state that two processes are responsible for the content of $^{210}$Pb: the $^{226}$Ra sub-decay chain, which contributes with about 11~mBq/kg, and an independent contamination of $^{210}$Pb which amounts to about 11~mBq/kg.

Given the high-radiopurity level of the archaeological Pb employed for the production of this crystal~\cite{Belli:2020}, the responsible for the residual contaminations can be ascribed mostly to the WO$_3$ used for the synthesis of the PbWO$_4$ powder. However, it is worth mentioning that the concentration of radionuclides inside the detector is by far the lowest ever measured in a PbWO$_4$ crystal. It is more than one order of magnitude better than the most sensitive detector reported in literature~\cite{Beeman:2012wz}. 

The high radiopurity level of this detector demonstrates the feasibility of producing large mass and low-background PbWO$_4$ crystal from archaeological Pb. The high quality of this valuable material can be preserved during the different steps of the crystal production, leading to a crystal with low radioactive contaminations. In addition, there is ample margin for improvement when highly radiopure WO$_3$ are used for the crystal production.
A detailed screening campaign for high purity WO$_3$ powder is presented in~\cite{2017PhDT}. There, samples from different producers were assayed, and some that can meet the requirements for the RES-NOVA crystal production are identified ($^{226}$Ra $<$2.2~mBq/kg).

\begin{table}
\caption{Evaluated internal radioactive contaminations for the PbWO$_4$ crystal. Nuclides of the decay chains with longer half-lives are listed. Limits are at 90\% C.L.} 
\begin{center}
\begin{tabular}{lcc}
\hline\noalign{\smallskip}
Chain & Nuclide  & Activity \\ 
            & & [mBq/kg] \\
\noalign{\smallskip}\hline\noalign{\smallskip}
$^{232}$Th & $^{232}$Th & $<$0.04 \\
 & $^{228}$Th & 0.80$\pm$0.09 \\
\noalign{\smallskip}\hline\noalign{\smallskip}
$^{238}$U & $^{238}$U & $<$0.03 \\
& $^{234}$U & $<$0.03 \\
& $^{230}$Th & $<$0.04 \\
& $^{226}$Ra & 11.34$\pm$0.35 \\
& $^{210}$Pb/$^{210}$Po & 22.50$\pm$0.49 \\
\noalign{\smallskip}\hline
\end{tabular}
\label{tab:radio} 

\end{center}
\end{table}

\section{Background extrapolation in the region of interest for SN neutrino detection}
\begin{figure}[h]
\centering
\includegraphics[width=0.49\textwidth]{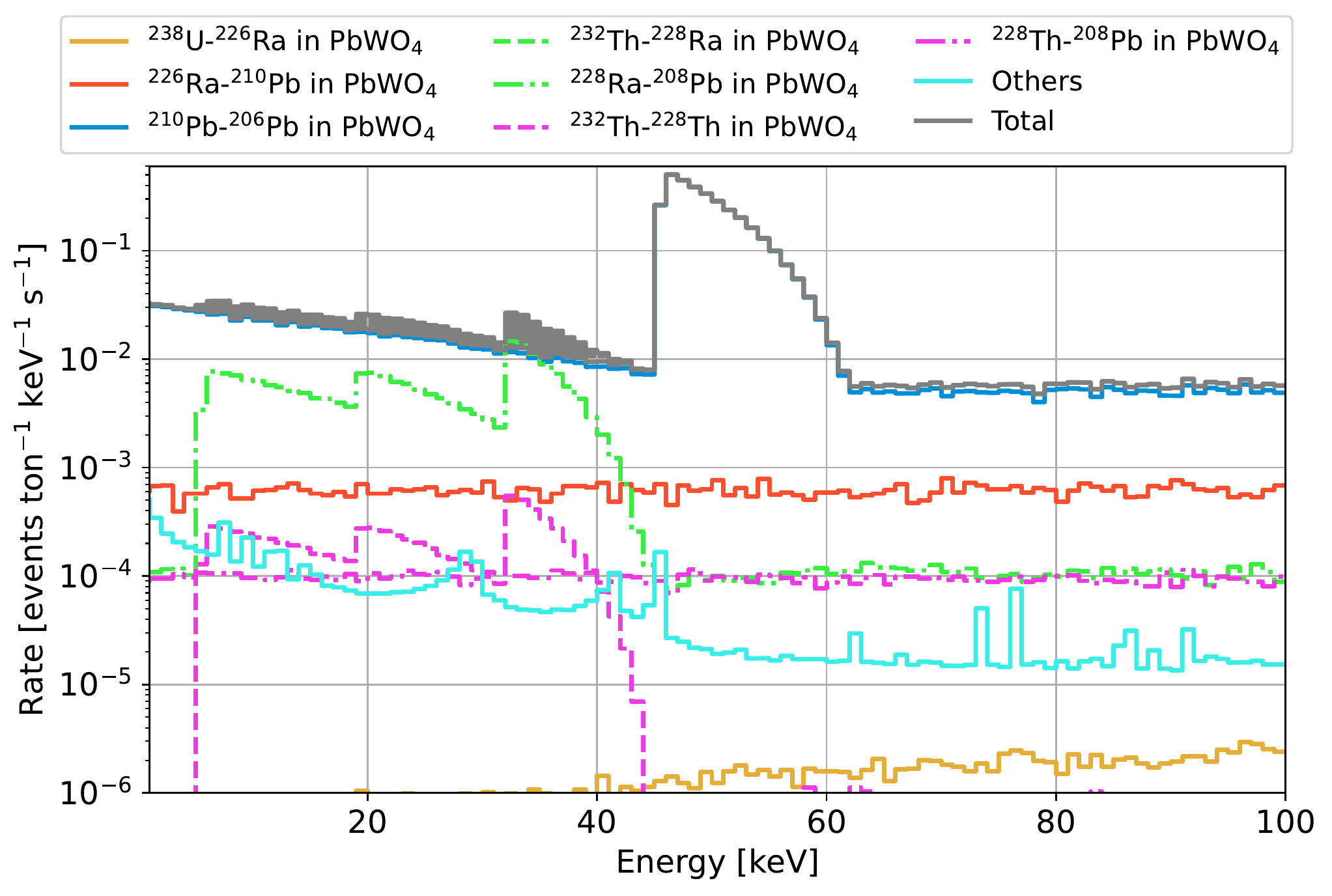}
\caption{RES-NOVA detector response to simulated radioactive contaminations listed in Tab.~\ref{tab:radio}. Different decay chains are simulated, taking into account the respective breaking point of equilibrium. The label \textit{Others} identifies all background sources not produced by the $^{arch}$PbWO$_4$ crystal. The gray band represents the sum of the different simulated energy spectra considering an $^{228}$Ac contamination ranging from $(0.04- 0.80)$~mBq/kg (for more details see the text). The energy spectra are zoomed in the region of interest for the detection of SN neutrinos via CE$\nu$NS.}
\label{fig:MC} 
\end{figure}

The important information extracted from the study on the radiopurity level of this large mass $^{arch}$PbWO$_4$ crystal provides the first opportunity for validating the RES-NOVA background model.
We used the results of our $\alpha$-analysis on the crystal internal contaminations as input to a Monte Carlo tool for the simulation of signals produced in the RES-NOVA detector by these contaminations. The Monte Carlo employed for this study was the same presented in~\cite{RES-NOVA:2021gqp}, where all known contaminations in the most relevant detector components were simulated (i.e. Cu structure, PTFE crystal holding system, Cu cryogenic vessels and a Polyethylene shield). The simulations generate the entire decay chain of the radioactive nuclides and also the breaking points of equilibrium. 
The outputs of the simulations are normalized for unit of primordial radioactive decay, and thus the activities of the crystal contaminations are used to scale the Monte Carlo outputs to the final energy spectrum values. The results are shown in Fig.~\ref{fig:MC}, where the contributions from the crystal and from all other background sources (labelled \textit{Others}) are presented. The colored band in the spectrum represents the expected minimum and maximum background level when the two different activities for $^{228}$Ac are considered, $(0.04-0.80)$~mBq/kg. We adopt a conservative approach and we consider the lower $^{228}$Ac ($^{232}$Th) activity limit as a value.
In addition, the Monte Carlo code takes into account the main detector parameters, such as the energy and time resolution, which were 0.2~keV and 100~$\mu$s respectively. The output produced is the final detector energy spectrum expected to be acquired by RES-NOVA during its background data taking.

The largest contribution to the background in the RoI, which lies between $[1,30]$~keV comes from $^{228}$Ac and $^{210}$Pb. While the latter has shown a clear contamination in the crystal, for the former a range of contamination needs to be taken into account $(0.04- 0.80)$~mBq/kg, as discussed in the previous section. In Fig.~\ref{fig:MC}, the total simulated energy spectrum expected to be observed in RES-NOVA is shown. This is the expected background level to be observed during the cooling phase of the SN neutrino emission, where the largest number of neutrinos is released.

When the total simulated energy spectrum is integrated over the RES-NOVA RoI for the detection of SN neutrinos, a background index ranging between (0.65-0.77)~events$\cdot$t$^{-1} \cdot$s$^{-1}$ is found. This background level is only a factor 9 to 11 away from the final goal of the experiment, and it is a factor $>$1000 better than the one achievable with a commercially available PbWO$_4$ crystal. RES-NOVA is aiming at achieving a background level in the 1-30~keV RoI for SN neutrino detection of about 0.001~c$\cdot$t$^{-1}$$\cdot$keV$^{-1}$$\cdot$s$^{-1}$, namely 0.002~c$\cdot$t$^{-1}$$\cdot$keV$^{-1}$$\cdot$s$^{-1}$ when the detector is operated in anti-coincidence mode~\cite{RES-NOVA:2021gqp}. While, as reported in \cite{RES-NOVA:2021gqp}, a conventional neutrino signal (27~$M_{\odot}$ progenitor star, model LS220) at 10~kpc would have induce a neutrino signal of 30 events over 10~s ($\sim$0.055~~c$\cdot$t$^{-1}$$\cdot$keV$^{-1}$$\cdot$s$^{-1}$).

It is worth mentioning that with the level of background extracted from the measurement reported in this paper, RES-NOVA would
still be sensitive to SN neutrinos, however with a reduced discovery potential (e.g. farther distances may not be surveyed)~\cite{RES-NOVA:2021gqp}.

\section{Outlook}
The radioactivity content of the crystal analysed in this work shows the feasibility in producing low-background PbWO$_4$ detectors, and that archaeological Pb can be employed for crystal production. However, the achieved results show that additional efforts are needed for reaching the background goal of the experiment. 
In particular, the selection of highly radiopure raw materials for the crystal production plays a crucial role, namely on WO$_3$. The background sources associated with WO$_3$ to be reduced are $^{210}$Pb and $^{228}$Ac. The first one can be reduced when the content of $^{226}$Ra in WO$_3$ is $<$1~mBq/kg. The second one requires a control of the $^{228}$Ra/$^{228}$Ac content, using an HP-Ge detector for $\gamma$-spectroscopy, to the level of $\sim 0.04$~mBq/kg.
We would like to mention also that, we are currently planning to investigate the radiopurity of another large mass $^{arch}$PbWO$_4$ crystal produced with more refined raw materials.

\section{Conclusions}
We presented the production of the first large mass PbWO$_4$ crystal produced from archaeological Pb, with a total mass of 0.84~kg. The crystal was successfully operated as a cryogenic particle detector in an underground experimental set-up. The crystal internal radiopurity level was assessed, nuclides from the $^{232}$Th decay chain and, from the lower part of the $^{238}$U one, were observed. A total $^{210}$Pb contamination of 15~mBq/kg was also measured. This crystal is by far the PbWO$_4$ with the highest radiopurity level ever measured. This important result was achieved thanks to the unique properties of archaeological Pb. 

The measured internal contaminations of the crystal were used for evaluating the background projection in the RoI for SN neutrino detection. This very first proof-of-principle detector allows reaching a background level which is a factor 9 to 11 away from the background goal of the RES-NOVA experiment. Mitigation actions can be easily implemented (e.g. selection of radiopure raw materials for the crystal production) for an effective background suppression. Future studies on the radiopurity of PbWO$_4$ crystals will require a strict control of $^{210}$Pb and $^{228}$Ac contaminations.

The operation of this new class of cryogenic detectors open a window of opportunities for the realization of a next-generation SN neutrino observatory with great physics potential.

\begin{acknowledgements}
We thank the CUPID collaboration for sharing their cryogenic infrastructure, M. Guetti for the assistance in the cryogenic operations, M. Perego for his invaluable help in many tasks, the mechanical workshop of LNGS.
This research was supported by the Excellence Cluster ORIGINS which is funded by the Deutsche Forschungsgemeinschaft (DFG, German Research Foundation) under Germany’s Excellence Strategy - EXC-2094 - 390783311.
F.A.~Danevich and V.I.~Tretyak were supported in part by the National Research Foundation of Ukraine Grant No. 2020.02/0011. This work makes use of the DIANA data analysis and APOLLO data acquisition software which has been developed by the CUORICINO, CUORE, LUCIFER and CUPID-0 collaborations. The authors would like to express their sorrow and regret about what happens in Ukraine.
\noindent The authors stand against the military actions initiated in Ukraine by the authorities of the Russian Federation.

\end{acknowledgements}

\bibliographystyle{spphys}       

\newcommand*{\doi}[1]{\href{https://doi.org/\detokenize{#1}}{doi: \detokenize{#1}}}

\bibliography{template}

\begin{thebibliography}{10}
\providecommand{\url}[1]{{#1}}
\providecommand{\urlprefix}{URL }
\expandafter\ifx\csname urlstyle\endcsname\relax
  \providecommand{\doi}[1]{DOI \discretionary{}{}{}#1}\else
  \providecommand{\doi}{DOI \discretionary{}{}{}\begingroup
  \urlstyle{rm}\Url}\fi

\bibitem{TXS_0506_056}
M.~Aartsen, et~al., Science \textbf{361}(6398) (2018).
\newblock \doi{10.1126/science.aat1378}

\bibitem{GW170817}
B.P. Abbott, et~al., Phys. Rev. Lett. \textbf{119}, 161101 (2017).
\newblock \doi{10.1103/PhysRevLett.119.161101}

\bibitem{PhysRevLett.58.1490}
K.~Hirata, et~al., Phys. Rev. Lett. \textbf{58}, 1490 (1987).
\newblock \doi{10.1103/PhysRevLett.58.1490}.
\newblock \urlprefix\url{https://link.aps.org/doi/10.1103/PhysRevLett.58.1490}

\bibitem{1987Nature}
R.~{Schaeffer}, et~al., Nat \textbf{330}(6144), 142 (1987).
\newblock \doi{10.1038/330142a0}

\bibitem{Mirizzi:2015eza}
A.~Mirizzi, I.~Tamborra, H.T. Janka, N.~Saviano, K.~Scholberg, R.~Bollig,
  L.~H{\"u}depohl, S.~Chakraborty, Riv. Nuovo Cim. \textbf{39}(1-2), 1 (2016).
\newblock \doi{10.1393/ncr/i2016-10120-8}

\bibitem{ALDUINO20199}
C.~Alduino, et~al., Cryogenics \textbf{102}, 9 (2019).
\newblock \doi{10.1016/j.cryogenics.2019.06.011}

\bibitem{Azzolini:2018tum}
O.~Azzolini, et~al., Eur. Phys. J. \textbf{C78}(5), 428 (2018).
\newblock \doi{10.1140/epjc/s10052-018-5896-8}

\bibitem{Cupid-Mo}
E.~Armengaud, et~al., Eur. Phys. J. C \textbf{80}(1), 44 (2020).
\newblock \doi{10.1140/epjc/s10052-019-7578-6}

\bibitem{Fink:2020jts}
C.W. Fink, et~al., Appl. Phys. Lett. \textbf{118}(2), 022601 (2021).
\newblock \doi{10.1063/5.0032372}

\bibitem{Alenkov:2019jis}
V.~Alenkov, et~al., Eur. Phys. J. C \textbf{79}(9), 791 (2019).
\newblock \doi{10.1140/epjc/s10052-019-7279-1}

\bibitem{Abdelhameed:2019hmk}
A.H. Abdelhameed, et~al., Phys. Rev. \textbf{D100}(10), 102002 (2019).
\newblock \doi{10.1103/PhysRevD.100.102002}

\bibitem{Strauss:2017cam}
R.~Strauss, et~al., Phys. Rev. D \textbf{96}(2), 022009 (2017).
\newblock \doi{10.1103/PhysRevD.96.022009}

\bibitem{Strauss:2017cuu}
R.~Strauss, et~al., Eur. Phys. J. C \textbf{77}, 506 (2017).
\newblock \doi{10.1140/epjc/s10052-017-5068-2}

\bibitem{Billard:2016giu}
J.~Billard, et~al., J. Phys. G \textbf{44}(10), 105101 (2017).
\newblock \doi{10.1088/1361-6471/aa83d0}

\bibitem{Aliane:2020dbc}
A.~Aliane, et~al., Nucl. Instrum. Meth. A \textbf{949}, 162784 (2020).
\newblock \doi{10.1016/j.nima.2019.162784}

\bibitem{Pattavina:2020cqc}
L.~Pattavina, N.~Ferreiro~Iachellini, I.~Tamborra, Phys. Rev. D
  \textbf{102}(6), 063001 (2020).
\newblock \doi{10.1103/PhysRevD.102.063001}

\bibitem{Freedman:1973yd}
D.Z. Freedman, Phys. Rev. D \textbf{9}, 1389 (1974).
\newblock \doi{10.1103/PhysRevD.9.1389}

\bibitem{RES-NOVA:2021gqp}
L.~Pattavina, et~al., JCAP \textbf{10}, 064 (2021).
\newblock \doi{10.1088/1475-7516/2021/10/064}

\bibitem{SNEWS:2020tbu}
S.~Al~Kharusi, et~al., New J. Phys. \textbf{23}(3), 031201 (2021).
\newblock \doi{10.1088/1367-2630/abde33}

\bibitem{Akimov:2017ade}
D.~Akimov, et~al., Science \textbf{357}(6356), 1123 (2017).
\newblock \doi{10.1126/science.aao0990}

\bibitem{Freedman:1977xn}
D.Z. Freedman, D.N. Schramm, D.L. Tubbs, Ann. Rev. Nucl. Part. Sci.
  \textbf{27}, 167 (1977).
\newblock \doi{10.1146/annurev.ns.27.120177.001123}

\bibitem{Prospects_Scholberg}
K.~Scholberg, Phys. Rev. D \textbf{73}, 033005 (2006).
\newblock \doi{10.1103/PhysRevD.73.033005}

\bibitem{Amanik:2009zz}
P.S. Amanik, G.C. McLaughlin, J. Phys. G \textbf{36}, 015105 (2009).
\newblock \doi{10.1088/0954-3899/36/1/015105}

\bibitem{Raj:2019sci}
N.~Raj, Phys. Rev. Lett. \textbf{124}(14), 141802 (2020).
\newblock \doi{10.1103/PhysRevLett.124.141802}

\bibitem{Lang:2016zhv}
R.F. Lang, C.~McCabe, S.~Reichard, M.~Selvi, I.~Tamborra, Phys. Rev. D
  \textbf{94}(10), 103009 (2016).
\newblock \doi{10.1103/PhysRevD.94.103009}

\bibitem{Agnes:2020pbw}
P.~Agnes, et~al., JCAP \textbf{03}, 043 (2021).
\newblock \doi{10.1088/1475-7516/2021/03/043}

\bibitem{Horowitz:2003cz}
C.J. Horowitz, K.J. Coakley, D.N. McKinsey, Phys. Rev. D \textbf{68}, 023005
  (2003).
\newblock \doi{10.1103/PhysRevD.68.023005}

\bibitem{Schumann:2019eaa}
M.~Schumann, J. Phys. G \textbf{46}(10), 103003 (2019).
\newblock \doi{10.1088/1361-6471/ab2ea5}

\bibitem{Pirro:2017ecr}
S.~Pirro, P.~Mauskopf, Ann. Rev. Nucl. Part. Sci. \textbf{67}, 161 (2017).
\newblock \doi{10.1146/annurev-nucl-101916-123130}

\bibitem{Beeman:2012wz}
J.W. Beeman, et~al., Eur. Phys. J. A \textbf{49}, 50 (2013).
\newblock \doi{10.1140/epja/i2013-13050-7}

\bibitem{Pattavina:2019pxw}
L.~Pattavina, et~al., Eur. Phys. J. A \textbf{55}, 127 (2019).
\newblock \doi{10.1140/epja/i2019-12809-0}

\bibitem{Iachellini:2021rmh}
N.~Iachellini~Ferreiro, et~al., in \emph{{19th International Workshop on Low
  Temperature Detectors}} (2021)

\bibitem{Abdelhameed:2020opm}
A.H. Abdelhameed, et~al., J. Low Temp. Phys. \textbf{199}(1-2), 401 (2020).
\newblock \doi{10.1007/s10909-020-02357-x}

\bibitem{Heusser:1995wd}
G.~Heusser, Ann. Rev. Nucl. Part. Sci. \textbf{45}, 543 (1995).
\newblock \doi{10.1146/annurev.ns.45.120195.002551}

\bibitem{Belli:2016}
P.~Belli, et~al., Phys. Rev. C \textbf{93}(4), 045502 (2016).
\newblock \doi{10.1103/PhysRevC.93.045502}

\bibitem{Belli:2020}
P.~Belli, et~al., Universe \textbf{6}(10), 182 (2020).
\newblock \doi{10.3390/universe6100182}

\bibitem{Danevich:2010}
F.A. Danevich, et~al., Nucl.\ Instrum.\ Meth.\ A \textbf{622}(3), 608 (2010).
\newblock \doi{https://doi.org/10.1016/j.nima.2010.07.060}

\bibitem{Alessandrello:1991}
A.~Alessandrello, et~al., Nucl.\ Instrum.\ Meth.\ B \textbf{61}(1), 106 (1991).
\newblock \doi{https://doi.org/10.1016/0168-583X(91)95569-Y}

\bibitem{Alessandrello:1993}
A.~Alessandrello, et~al., Nucl.\ Instrum.\ Meth.\ B \textbf{83}, 539 (1993).
\newblock \doi{10.1016/0168-583X(93)95884-8}

\bibitem{Danevich:2009}
F.A. Danevich, et~al., Nucl.\ Instrum.\ Meth.\ A \textbf{603}(3), 328 (2009).
\newblock \doi{https://doi.org/10.1016/j.nima.2009.02.018}

\bibitem{Boiko:2011}
R.S. Boiko, et~al., Inorganic Materials \textbf{47}(6), 645 (2011).
\newblock \doi{10.1134/S0020168511060069}

\bibitem{Cardani:2013dia}
L.~Cardani, et~al., JINST \textbf{8}, P10002 (2013).
\newblock \doi{10.1088/1748-0221/8/10/P10002}

\bibitem{Casali:2013zzr}
N.~Casali, et~al., J. Phys. \textbf{G41}, 075101 (2014).
\newblock \doi{10.1088/0954-3899/41/7/075101}

\bibitem{Pattavina:2015jxe}
L.~Pattavina, et~al., J. Low Temp. Phys. \textbf{184}(1-2), 286 (2016).
\newblock \doi{10.1007/s10909-015-1404-9}

\bibitem{Artusa:2016mat}
D.R. Artusa, et~al., Phys. Lett. B \textbf{767}, 321 (2017).
\newblock \doi{10.1016/j.physletb.2017.02.011}

\bibitem{JFET}
C.~Arnaboldi, et~al., JINST \textbf{13}(02), P02026 (2018).
\newblock \doi{10.1088/1748-0221/13/02/p02026}

\bibitem{ARNABOLDI2010327}
C.~Arnaboldi, et~al., Nucl. Instrum. Meth. A \textbf{617}(1), 327 (2010).
\newblock \doi{https://doi.org/10.1016/j.nima.2009.09.023}

\bibitem{Domizio_2018}
S.D. Domizio, A.~Branca, A.~Caminata, L.~Canonica, S.~Copello, A.~Giachero,
  E.~Guardincerri, L.~Marini, M.~Pallavicini, M.~Vignati, Journal of
  Instrumentation \textbf{13}(12), P12003 (2018).
\newblock \doi{10.1088/1748-0221/13/12/p12003}

\bibitem{Gatti:1986cw}
E.~Gatti, P.F. Manfredi, Riv. Nuovo Cim. \textbf{9N1}, 1 (1986).
\newblock \doi{10.1007/BF02822156}

\bibitem{Azzolini:2018yye}
O.~Azzolini, et~al., Eur. Phys. J. C \textbf{78}(9), 734 (2018).
\newblock \doi{10.1140/epjc/s10052-018-6202-5}

\bibitem{Andreotti:2010vj}
E.~Andreotti, et~al., Astropart. Phys. \textbf{34}, 822 (2011).
\newblock \doi{10.1016/j.astropartphys.2011.02.002}

\bibitem{Alfonso:2015wka}
K.~Alfonso, et~al., Phys. Rev. Lett. \textbf{115}(10), 102502 (2015).
\newblock \doi{10.1103/PhysRevLett.115.102502}

\bibitem{Clemenza:2011zz}
M.~Clemenza, C.~Maiano, L.~Pattavina, E.~Previtali, Eur. Phys. J. C
  \textbf{71}, 1805 (2011).
\newblock \doi{10.1140/epjc/s10052-011-1805-0}

\bibitem{Lecoq:1994yr}
P.~Lecoq, et~al., Nucl. Instrum. Meth. A \textbf{365}, 291 (1995).
\newblock \doi{10.1016/0168-9002(95)00589-7}

\bibitem{Borexino:2012uda}
G.~Bellini, et~al., Eur. Phys. J. A \textbf{49}, 92 (2013).
\newblock \doi{10.1140/epja/i2013-13092-9}

\bibitem{LYS}
M.~Clemenza, G.~Cucciati, V.~Maggi, L.~Pattavina, E.~Previtali, Eur. Phys. J.
  Plus \textbf{127}(6), 68 (2012).
\newblock \doi{10.1140/epjp/i2012-12068-0}

\bibitem{Strauss:2014aqw}
R.~Strauss, et~al., JCAP \textbf{06}, 030 (2015).
\newblock \doi{10.1088/1475-7516/2015/06/030}

\bibitem{2017PhDT}
A.R. {M{\"u}nster}, {High-purity CaWO$_{4}$ single crystals for direct dark
  matter search with the CRESST experiment}.
\newblock Ph.D. thesis, Technical University of Munich, Germany (2017)

\end{thebibliography}

\end{document}